# Stern Layers on Surfaces of $RuO_2(100)$, $RuO_2(110)$, and Pt(111): Surface X-ray Scattering Studies

Tomoya Kawaguchi[a,†], Reshma R. Rao[b], Jaclyn R. Lunger[c], Yihua Liu[a,§], Donald A. Walko[d], Evguenia A. Karapetrova[d], Vladimir Komanicky[e], Yang Shao-Horn[b,c], Hoydoo You[a,*]

[a] *Materials Science Division, Argonne National Laboratory, Argonne, Illinois, 60439, United States*

[b] *Department of Mechanical Engineering, Massachusetts Institute of Technology, Cambridge, Massachusetts 02139, United States*

[c] *Department of Materials Science and Engineering, Massachusetts Institute of Technology, Cambridge, Massachusetts 02139, United States*

[d] *X-ray Science Division, Argonne National Laboratory, Argonne, Illinois, 60439, United States*

[e] *Safarik University, Faculty of Sciences, Kosice 04154, Slovakia*

\* *hyou@anl.gov*

[†] Current Address: Institute for Materials Research, Tohoku University, Japan

[§] Current Address: Lam Research, Tualatin, OR, USA

**Abstract**

Surface X-ray scattering studies of electrochemical Stern layer are reported. The Stern layers formed at the interfaces of $RuO_2$ (110) and (100) in 0.1 M CsF electrolyte are compared to the previously reported Stern layer on Pt(111) [Liu et al., J. Phys. Chem. Lett., 9 (2018) 1265]. While the $Cs^+$ density profiles at the potentials close to hydrogen evolution reaction are similar, the hydration layers intervening the surface and the $Cs^+$ layer on $RuO_2$ surfaces are significantly denser than the hydration layer on Pt(111) surface possibly due to the oxygen termination of $RuO_2$ surfaces. We also discuss in-plane ordering in the Stern layer on Pt(111) surface.

## 1. Introduction

The structure and motion of ions in electrochemical double layers (EDL) is key information in basic surface electrochemistry [1]. EDL is traditionally described as diffuse distributions of cations and anions, proposed by Gouy and Chapman [2] and further developed extensively over decades [3-10] and the early development of the diffuse EDL models can be found in a review [11]. The early studies are mainly based on the voltammetry measurements without direct structural studies. In recent years, synchrotron X-ray techniques have been widely used for studies of chemical and electrochemical double layers on various interfaces such as the membrane-aqueous interfaces using X-ray standing wave technique [12], the solid/liquid interfaces using crystal truncation rod measurements [13], and liquid/liquid interfaces using X-ray reflectivity techniques[14].

While it is generally accepted that Gouy-Chapman model or its modified versions describe well electrochemical interfacial phenomena, our recent study of Pt(111) surface in CsF solution has shown that Stern layer forms over a large portion of the double-layer potential range [15] using a model-independent direct inversion method [16]. In this study [15], the formation of Stern layer was additionally supported in the time-dependent recoil behavior of the top Pt layer in potential-jump experiments in various alkali solutions.

In this report, a parallel study of EDL on $RuO_2$ surfaces in CsF solution will be presented. $RuO_2$ is chosen because it is one of the most active electrocatalytic materials in oxygen evolution reactions [17-25] and also extensively studied for hydrogen evolution reactions [18, 19, 26-29]. For the reason, the surface structures are well known from early X-ray studies [27, 30, 31]. In addition, the surface symmetries of $RuO_2$ are either square or rectangular while the symmetry of Pt(111) is triangular. More importantly, the chemistries of the terminating surfaces are expected to be very different between oxides and metals. Therefore, it is of interest to test and compare the structures and distributions of EDL formed on RuO2 to those on Pt(111).

## 2. Experimental

### 2.1 Transmission Cell

An transmission cell, similar to a drop cell [32] and improved from the previous transmission cells [33, 34], was used for this experiment. In this cell, the sample surface can be cleaned, annealed, and immersed to an electrolyte without exposing the surface to ambient conditions. In this way, the pristine as-prepared surface is immediately used for electrochemical X-ray measurements. The cell geometries are schematically shown in Figure 1. In (a), X rays diffract from the surface through the electrolyte typically ~2 mm thick. X-ray transmission tends to produce background scattering. In (b), X-rays do not go through the water, therefore, the background scattering is lower. However, the electrochemical control is lost and the situation is equivalent to the *ex situ* condition in the UHV transfer experiments [35]. In (c), the inner quartz pipet, assembled with reference and counter electrodes, is fully retracted and the induction heater coil is moved to align with the sample height. This configuration is used for sample annealing of Pt(111) surface in 3%H$_2$/Ar flow and surface cleaning of RuO$_2$ in O$_2$ flow.

### 2.2 Sample Preparations

The RuO$_2$ crystals were grown in a tube furnace using the oxidative sublimation/crystallization method [31]. The most abundant and largest crystal faces typically of a few mm in size were the (110) and (100) orientations. In most cases, the single crystals grown this way exhibited excellent quality of bulk mosaics with few defects such as twinning or faults. A typical bulk mosaic was ~0.01° close to the quality of a perfect single crystal. Pt(111) surface was prepared from commercially available single crystals. The crystal was precut, polished, and annealed [36] until the mosaic and the miscut of the crystal became < 0.1°. For electrochemical measurements, 0.1 M CsF electrolytes was prepared from the CsF solid with a purity of 99.99% from Puratronic® dissolved in 18 MΩ·cm water.

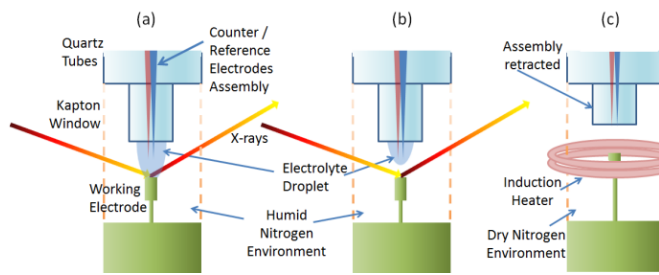

Figure 1. Schematics of the transmission cell: (a) The working electrode surface is immersed for electrochemical control and *in situ* transmission x-ray measurements. (b) The electrolyte droplet is lifted for *ex situ* measurements. (c) The counter/reference electrodes assembly is retracted and the electrode assembly was raised for the inductive annealing of the electrode.

## 2.3 Synchrotron X-ray Measurements

Synchrotron X-ray measurements were performed at 33BM, 7ID, and 11ID-D beamlines equipped with a '4S+2D' geometry six-circle diffractometer [37] at Advanced Photon Source (APS). For $RuO_2$ surfaces, the natural indices are used for CTR measurements. For (100) surface, one unit cell include two ½ Ru layers and four ½ oxygen layers. For (110) surface, one unit cell include two ½ Ru-O layers and four ¼ oxygen layers. CTR measurements were performed in the unit of H in both cases. For Pt(111) surface, the hexagonal (hex) index ($a^* = 4\pi\sqrt{2} / \sqrt{3}a$ and $c^* = 2\pi / \sqrt{3}a$ where a=3.9242Å) instead of face-centered cubic (fcc) indices was used in the experiments [38] where $(111)_{fcc}$, $(1\bar{1}1)_{fcc}$, and $(200)_{fcc}$ are indexed to $(003)_{hex}$, $(101)_{hex}$, and $(012)_{hex}$, respectively. The surfaces were checked for readiness and pre-oriented by X-ray reflectivity and Bragg diffraction before made in contact with the electrolyte droplet. In addition, the X-ray shutter was open only during active photon counting and experiments were often repeated with attenuated intensities of incoming X-rays to ensure that the results are in any way affected by the degree of X-ray exposure because open circuit potentials can drift during X-ray exposure [39].

## 3. Results and Discussion

X-ray Crystal truncation rods [40, 41] (CTR) was used in this study. CTR is a powerful X-ray technique for electrode surface structures [42, 43]. Its high sensitivity was used to determine the structure of light molecules on metallic electrodes, such as layering of water molecules [30, 44]. The sensitivity to light elements can be further enhanced by modeling the CTR normalized by a 'standard' CTR [36] because the normalization reduce the effect of the insensitive but dominant near-Bragg intensities on the overall fits. In this way, the normalized CTRs are sensitive even to a weak density perturbation such as EDL profiles. The normalized CTR, $I_{norm}$, is defined as,

$$I_{norm} = \frac{|F_{sub}+F_{electrolyte}+F_{expansion}+F_{EDL}|^2}{|F_{sub}+F_{electrolyte}|^2} \tag{1}$$

where $F_{sub}$ is the CTR structure factor of the standard state, $F_{electrolyte}$ is the scattering factor of the semi-infinite electrolyte density (mainly water), $F_{expansion}$ is the contribution to CTR by the expansion or contraction of the top crystal surface layers, and $F_{EDL}$ is the scattering contribution from the electrochemical double layers composed of cation, $Cs^+$ and hydration layers. Here $F_{sub}$ + $F_{electrolyte}$ is the structure factor of the standard CTR. The standard state may include lattice expansions and EDL modulations even though the standard potential was chosen for the least perturbed substrate and EDL density profile. For the reason, the obtained density profiles are the *differences* or deviations from the standard EDL profile, which may not accurately represent the real EDL profiles if the standard state is not completely known. The equations used in modeling the EDL structures for $RuO_2$ surfaces are given in the supplementary information [45] and the model for Pt(111) surface is given in the supplementary information of Ref. [15]. In this report, we included the water density explicitly in the standard state to provide the overall density modulations with respect to the background electron density of water, 0.33 $e^-/Å^3$. The normalized data were fit to the four peak model used in the previous EDL study on Pt(111) surface [15], which was derived from the direct inversion method [16].

## 3.1 RuO$_2$ (110) and (100) Surfaces

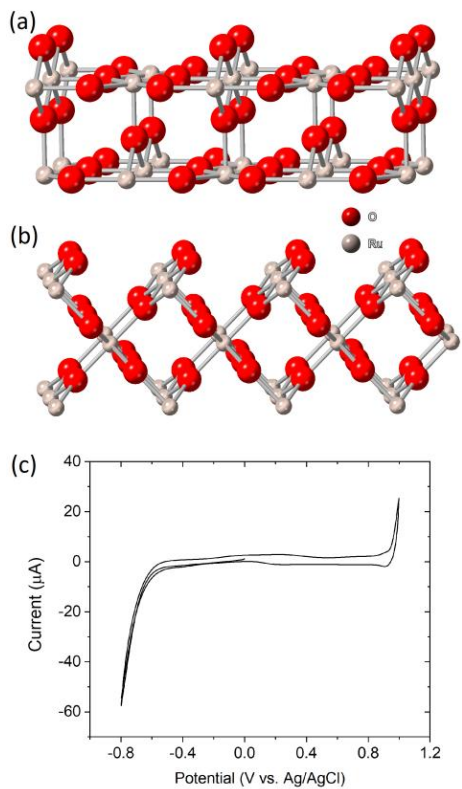

The surface structures of RuO$_2$ are shown in Figure 2. In (a), the (110) surface is shown with the top oxygen layer bridging the Ru atoms of the top Ru-O layer. The surface unit cell of rutile structure in this orientation can be defined by six layers; the top Ru-O layer, two ½-occupancy oxygen layers, the staggered second Ru-O layer, and two more oxygen layers. The two ½-occupancy oxygen layers below the second Ru-O layer are not shown. Similarly, in (b), the (100) surface is shown with the top-layer oxygen atoms bridging Ru atoms, Ru layer with two oxygen layers, the staggered second Ru-O layer with two oxygen layers, and finally with the 3$^{rd}$ Ru layer. In (c), the cyclic voltammogram (CV) for (110) surface is shown. The redox features at ~0.3 V are the peaks associated with adding/removing atop oxygen [30] without significantly disturbing the bridging oxygen layer. The CV for (100) surface is similar.

Figure 2. Surface structures of RuO$_2$ (110) and (100) shown with a terminating bridging oxygen atoms. (c) CV for the (110) surface in 0.1 M CsF obtained in the transmission cell.

The standard potential was chosen so that there is no structure changes for cathodic scans from the potential. The potential range (−1000 mV to 0 mV) are chosen so that no addition or removal of oxygen bonds on the surface. CTRs for RuO$_2$(110) surface were measured at 0, −200, −500, −700, −1000 mV vs. Ag/AgCl in 0.1 M CsF solution. Then, the CTRs were normalized by the 0 mV CTR. The normalized CTRs for −200, −500, −700, −1000 mV and the corresponding fits are shown in Figure 3(a). In (b), the density profile obtained from the fits for −200, −500, −700 mV are shown. The density profile for −1000 mV is not shown because it is similar to that for −700 mV. Despite the imperfect surfaces of as-grown $RuO_2$ crystals, main features of the density profiles are comparable to the results for Pt(111) surface. In (c) the previous results for Pt(111) surface [15] are shown with the water background (0.33 $e^-/Å^3$) for direct comparison. The standard potential for Pt(111) was +400 mV. The solid black lines in (b) and (c) indicate the approximate water density profiles.

In comparing (b) and (c), the distance of the main $Cs^+$ peak at ~4 Å from $RuO_2$(110) surface is similar to the distance of the $Cs^+$ peak from Pt(111) surface. In particular, the profiles for −700 mV in (b) and −850 mV in (c) are quite similar, suggesting that EDLs are not very sensitive to the details of electrode structures or symmetries. However, there are also some differences. (i) The first peak at ~1 Å for $RuO_2$(110) is significantly larger than that for Pt(111). This is possibly because the terminating bridging oxygen atoms and the coordinatively unsaturated (CUS) Ru atoms on the surface can hold many more water molecules via hydrogen bonds [30] than the bare platinum atoms on Pt(111). (ii) The water density and the first peak appear penetrating the top

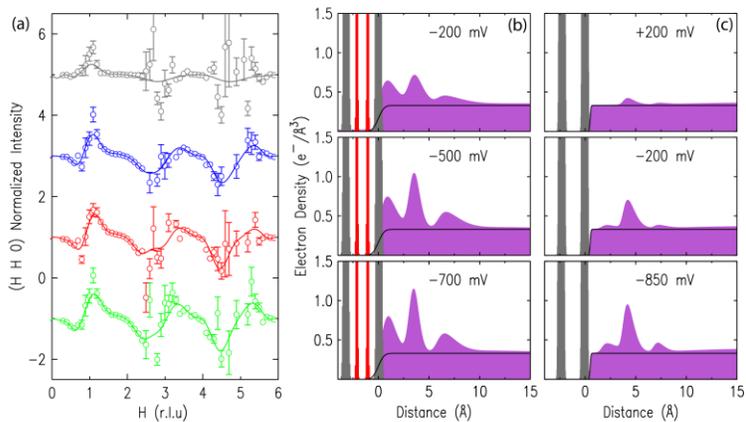

Figure 3. (a) Fits to normalized CTR data, from the top, at −200, −500, −700, −1000 mV vs. Ag/AgCl in 0.1 M CsF solution. (b) The density profiles obtained from the fits at −200, −500, and −700 mV and (c) the results from Pt(111) surface are shown together for comparision. The solid black lines indicate the electrolyte density. The vertical lines in (b) indicate the surface Ru-O layer (grey) and O (red) layer and the grey vertical lines in (c) does the Pt layer positions.

Ru-O layer. The apparent overlap with the surface layers is the effect of surface steps. The interface are broadened due to surface steps on the as-grown $RuO_2$ single crystal surfaces [31]. The surface with steps effectively broadens the profiles of surface layers and make them appear overlapping. On the contrary, there is almost no overlap in the Pt(111) case because Pt(111) surfaces were prepared to a near step-free perfection by using high-temperature annealing.

There are significant differences between −200 mV in (b) and +200 mV in (c). It is surprising because both of them are only 200 mV away from the standard potentials. On closer examination, −500 mV profile is almost as strong as that for −700 mV for $RuO_2$ while −200 mV and −850 mV profiles for Pt(111) increase gradually. These observations indicate Stern layer formation is considerably more abrupt at $RuO_2$(110) surface than Pt(111) surface. This is probably again due to the terminating bridging oxygen atoms on the surface, which can draw water molecules via hydrogen bonds and the so polarized water molecules can in turn attract $Cs^+$ ions more readily at smaller overpotentials than the bare platinum atoms on Pt(111) surface can attract water and $Cs^+$ ions.

The results from $RuO_2$(100) surface is shown in Figure 4. The results are essentially the same as those for $RuO_2$(110), again confirming that EDL profiles are insensitive to the surface structures or symmetries. In fact, all other aspects of the EDL characteristics are similar to the (110) surface. The peak at ~1 Å overlaps with the top Ru layer and significantly stronger compared to the Pt(111) EDL profile, −200 mV profile show significantly larger peaks compared to Pt(111) at +200 mV, the profiles of $RuO_2$(100) at −500 mV and −700 mV are similar indicating the faster onset of the EDL profiles compared to Pt(111) surface. The EDL peaks are stronger than those of (110) surface probably because the (100) surface has a more open structure than (110) surface and can polarize more water molecules. Both (100) and (110) surface have the stronger EDL peaks compared to those of Pt(111) again because of the terminating oxygen atoms and CUS Ru sites attracting and polarizing water layers and subsequently attracting $Cs^+$ ions to the $RuO_2$ surfaces.

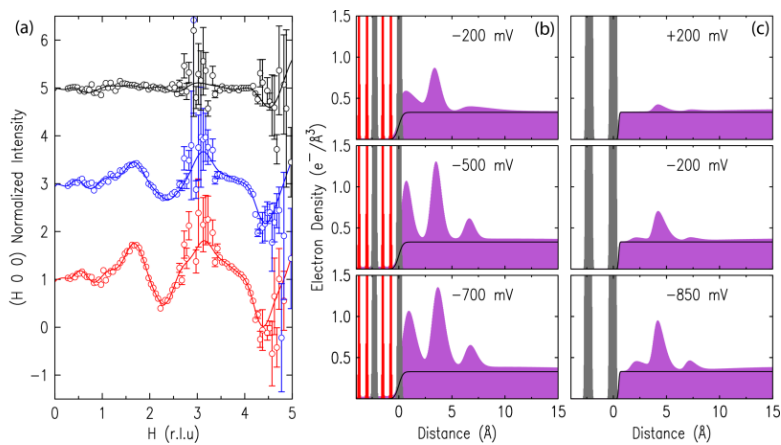

Figure 4. (a) Fits to the normalized CTR for $RuO_2$(100) surface, from the top, at −200 mV, −500 mV, and −700 mV, in 0.1 M CsF solution. (b) EDL density distributions. The vertical lines indicate the surface Ru (grey) and O (red) atomic position. (c) Pt(111) results. The vertical lines in (b) indicate the surface Ru layer (grey) and O (red) layer and the grey vertical lines in (c) does the Pt layer positions.

### 3.2 Search for in-plane EDL structure of $RuO_2$ surfaces

In-plane scans were made *in situ* while the potential was held at −700 mV to search for in-plane peaks on both (100) and (110) surfaces in electrolyte. However, we found no in-plane peaks probably because there is no long-ranged in-plane structure formation on $RuO_2$ surfaces. In-plane scans were repeated *ex situ* after emersion of the surfaces by lifting the electrolyte droplets. In the *ex situ* case, we found several in-plane peaks. However, none of them were commensurate or epitaxially oriented. Therefore, we concluded that the in-plane structure of Stern layer *is* affected by the structure and symmetry of the electrode surface. This observation can be compared to the triangular commensurate in-plane structures found *in situ* and *ex situ* in the case of Pt(111) surface. The details of the in-plane structure studies on Pt(111) will be discussed in the next section.

### 3.3 In-plane EDL structure on Pt(111) Surface

*In situ* search for in-plane peaks was performed first. While the potential was held at −850 mV, various commensurate in-plane reciprocal positions were scanned until a weak surface peak at (0.5 0.5 0.33) was identified. Then, the electrolyte droplet was withdrawn from the surface (see Figure 1b) at −850 mV while holding the surface vertical for a quick emersion. The scan through (0.5 0.5 0.33) was immediately repeated and several successive scans are shown in Figure 5. The integrated intensities measured *in situ* and measured *ex situ* at 1 min, at 6 min, and at 64 min after the emersion were 0.9(1), 2.2(1), 8.9(1), and 8.5(1), respectively. The scan at 0.4 V was flat within the noise and was used for the background subtraction. The intensity grows rapidly for the first 6 min after the emersion. It is important to note that the peak intensity goes back to that of the *in situ* condition if the surface is re-immersed immediately after the 1 min measurements. This indicates that the $Cs^+$ remains hydrated up to this point. If the surface is re-immersed after several min, however, the intensity remains strong and unresponsive to the applied potential, indicating that $Cs^+$ is dehydrated, at least partially, and possibly chemisorbed (or strongly adsorbed) to Pt(111) surface. In this case, the surface has to be reannealed to recover a clean surface. The intensity decreases eventually even under the humid $N_2$ flow in an hour after the emersion. The surface is no longer clean, probably due to the oxygen impurities in the cell. At this point, again, the surface has to be re-prepared to recover the pristine surface condition.

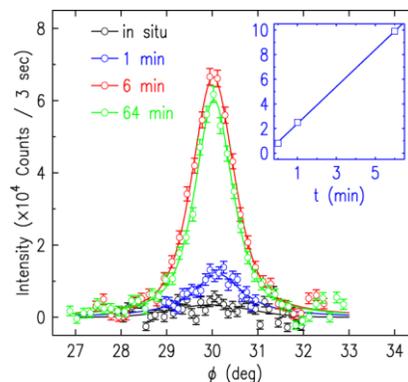

Figure 5. Scans at (0.5 0.5 0.33) *in situ* and *ex situ* at 1, 6, and 64 min after the emersion. The inset shows the integrated intensities mesasured at 0 (*in situ*), 1, and 6 min.

The superlattice peak at (0.5 0.5 0.33) indicates a (2×2) structure. In order to determine the structure, scans through three (2×2) peaks, (1 −0.5 0.33), (0.5 0 0.33), and (0.5 0.5 0.33), were made. A scan through Pt (0 1 0.5) was also made as a calibration point. These peaks are compared in Figure 6. The inset shows the reciprocal space map with the curved green arrows indicating the directions of the displayed transverse $\phi$ scans. Comparing the widths, the superlattice peaks are ~20 times broader than the Pt (0 1 0.5) peak. The longitudinal scans (not shown) are also ~10 times broader. These scans indicate that the average domain size of the superlattice is in the order of ~30 nm. The intensities of these peaks all show little dependence on *L* values, indicating that they are indeed from a monolayer structure. Pt (0 1 0.5) is the anti-Bragg peak between two Bragg peaks, (0 1 2) and (0 1 −1) and the intensity of this peak can be estimated from an expression, $\left|\frac{f_{Pt}}{1-e^{i2\pi(L-1)}}\right|^2$ for (0 1 *L*) crystal truncation rod [40], where $f_{Pt}$ is the form factor of a platinum atom. The form factor at small diffraction angles is essentially the atomic numbers ($f_{Pt}$ = 78). The calculated intensity of (0 1 0.5) is then $[4(78/2)]^2$ for a (2×2) unit cell where 4 comes from 4 atoms in the unit cell.

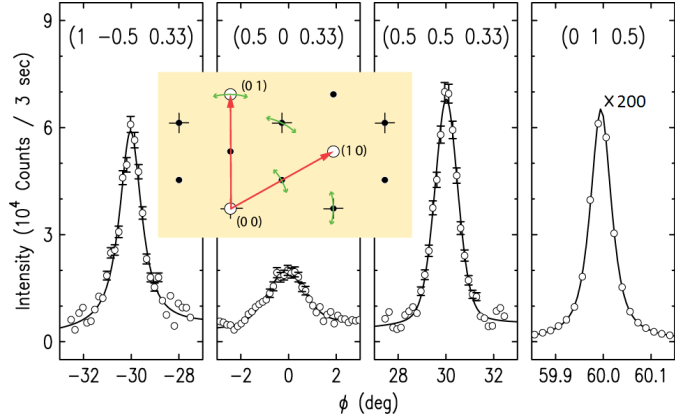

Figure 6. Three $Cs^+$ peaks and a Pt surface peak. (1 −0.5 0.33), (0.5 0 0.33), and (0.5 0.5 0.33) are the superlattice peaks due to the (2×2) $Cs^+$ layer and (0 1 0.5) is a Pt surface anti-Bragg peak. The inset shows a 2d in-plane reciprocal space. The green curved arrows indicate the directions of the scans.

In calculating the (2×2) superlattice peaks, three models (Figure 7) are considered. Other models beside them can be considered as variations or combinations of them by moving the $Cs^+$ positions to non-symmetric sites. Therefore, only these three models will be considered; two sublattice model (a), three sublattice model (b), and a single sublattice model (c). Since $Cs^+$ ions are hydrated, water molecules are expected to be incorporated into the lattice. However, they are not included in the calculations because they are weak scatters. The calculated intensities for the three models and the experimental intensities are compared in Table 1. The all values are scaled by setting the integrated intensity of (0 1 0.5) as unity.

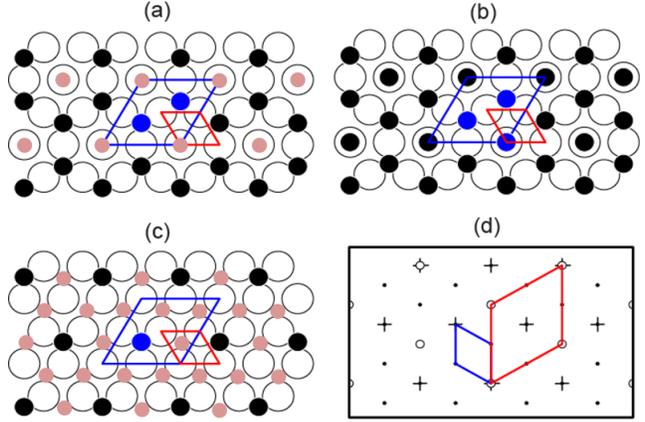

Figure 7. Three (2×2) models considered: (a) two sublattice unitcell, (b) three sublattice unitcell, (c) single sublattice unitcell. The open circles indicate Pt atoms, black solid circles represents Cs+ ions, and small pink circles show possible sites for water molecules. (d) shows the receiprocal space unitcells for Pt (1×1) (red) and Cs$^+$ (2×2) (blue). The circles are Pt receiprocal lattice and + and dots are the (2×2) superlattices.

Table 1. The comparison of the calculated (2×2) intensities to the measured *ex situ* and *in situ* intensities for the models shown in Figure 7. The *in situ* intensities are measured in 1 min after emersion. All intensities are normalized by (0 1 0.5) intensity.

|  | (a) | (b) | (c) | *ex situ* intensity | *in situ* intensity |
|---|---|---|---|---|---|
| (0.5 0.5 0.33) | 0.50 | 1.12 | 0.09 | 0.10(1) | 0.03(1) |
| (1. −0.5 0.33) | 0.50 | 1.12 | 0.09 | 0.9(1) | 0.03(1) |
| (0.5 0.0 0.33) | 0.12 | 0 | 0.09 | 0.03(1) | 0.03(1) |

The experimental integrated intensities in Table 1 are from the peaks shown in Figure 6. Let us consider the *ex situ* measurements first. The intensities for (0.5 0.5 0.33) and (1 −0.5 0.33) are 10% and 9%, respectively, and that for (0.5 0 0.33) is 3%. The Debye-Waller (DW) factors of Cs$^+$, which are unknown, are not included. Therefore, the calculated intensities are the upper bounds and the experimental intensities must be similar to the calculated values, if the Cs$^+$ layer is as well ordered as Pt layer. Otherwise, the intensity will smaller than the calculated ones because of DW factor. This eliminates the single sublattice model (c). Among (a) and (b), (b) can easily eliminated because the measured intensity of (0.5 0 0.33) is small but not zero. In the case of model (a), the intensities make sense if the DW factor significantly reduced all of the in-plane intensities. The Cs$^+$ ions are expected to be quite disordered with significant domain boundaries because Cs$^+$ domain sizes are much smaller than the Pt(111) surface domain size as discussed earlier. The theoretical DW factor is $e^{-(\sigma q)^2/3}$, where $q=4\pi\sin(\theta)/\lambda$ and $\sigma$ is the mean displacement. The 80% reduction by the DW factor suggests $\sigma$ = ~0.9 Å, which is almost ~33 % of the Pt-Pt distance. This is not unreasonable when the Cs+ are probably fully or partially hydrated and adsorbed to the surface with the water molecules. The measured intensity ratio between (0.5 0.5 0.33) and (0.5 0 0.33) also agrees with the calculated ratio with the DW factor.

For the *in situ* measurements, the (0.5 0.5 0.33) intensity (Figure 5) is barely above the background. The *in situ* intensity of (0.5 0 0.33) is also close to the background (not shown). In the first scan immediately after emersion, the (0.5 0.5 0.33) intensity is ~3% of the (0 1 0.5) intensity or ~25% of the full *ex situ* intensity. The scan time for the first scan is about a minute, which includes the time for withdrawing the electrolyte, interlocking the door of the X-ray hutch, and scanning the peak. Likewise, the (0.5 0 0.33) intensity, measured immediately following the first (0.5 0.5 0.33) measurement (about ~2 min after the emersion), is again ~3% of the (0 1 0.5) intensity. Note that the *ex situ* (0.5 0 0.33) intensity does not change over time while the *ex situ* (0.5 0.5 0.33) and (1 −0.5 0.33) peaks grow in time (Figure 5). The integrated intensities measured within 2 min indicate that the structures are different between the *ex situ* and *in situ* conditions. It suggests that additional Cs$^+$ ions must be incorporated into the (2×2) structure during the first several minutes after emersion, probably from the thin electrolyte layer invisible yet still remaining after the emersion. Therefore, the *in situ* (2×2) structure should be close to the model (c) where the intensities of (0.5 0.5 0.33), (1 −0.5 0.33), and (0.5 0 0.33) are all weak and similar each other. Since the calculated intensities of the model (c) are 9% each, the experimental observed value of ~3% each is reasonable for the model (c) considering the DW factor. This is also consistent with the CTR measurements [15, 16] where the direct inversion technique was used to obtain the $Cs^+$ layer density of ~0.7 $e^-/Å^3$, which is ~25% of 2.7 $e^-/Å^3$ for Pt layer electron density. Under this scenario, the highly disordered hydrated $Cs^+$ ions maintain the short-range single-sublattice (2×2) model (c) structure in electrolyte. As the solution thins in several min after emersion, however, additional $Cs^+$ ions are incorporated into the lattice to form the two-sublattice model (a) structure.

In summary, the charged ions, sometimes hydrated, can be strongly pulled toward the solid surface and form relatively dense layers of the ions for potentials significantly away from the potential of zero charge (PZC). The dense layer of the ions with repulsive interactions due to the same charges may induce significant in-plane structures that have not been observed heretofore. This in-plane structure, the focus of this study, differs from those occurring in chemisorption that must accompany faradaic charge transfer reactions. In the case of $Cs^+$ studied here, the cations do not specifically chemisorb even at the largest negative potential that we measured *in situ*, which is clear from the potential jump desorption experiments [15]. Yet, in-plane ordering peaks are identified under *in situ* as well as *ex situ* [35] conditions using the emersion technique[46-48].

## 4. Conclusions

Despite the differences in surface symmetries of $RuO_2(100)$ and $RuO_2(110)$ and $Pt(111)$ surfaces, the EDL distributions normal to the interface are similar, suggesting that Stern layer formation is probably universal regardless of surface structure or electrode materials. Stern layer formed the $Cs^+$ layer and two water layers; a water layer between the $Cs^+$ layer and the substrate and another water layer above for hydrating the $Cs^+$ ions. Finally a diffuse layer similar to the diffuse double layer predicted by Gouy-Chapman model exists with a significant distance (~10 Å) from the surface. While no in-plane ordering of $Cs^+$ ions on $RuO_2$ surfaces was observed, (2×2) peaks were observed in the *in situ* and *ex situ* measurements on Pt(111). The *in situ* (0.5 0.5) peak intensity was weak and consistent with the single-sublattice (2×2) structure and *ex situ* (0.5 0.5) peak was strong and consistent with two sublattice (2×2) structure.

**Acknowledgements**

The X-ray and electrochemistry work and data analysis were supported by the U.S. Department of Energy (DOE), Office of Basic Energy Science (BES), Materials Sciences and Engineering Division and use of the APS was supported by DOE BES Scientific User Facilities Division, under Contract No. DE-AC02-06CH11357. The work (VK) at Safarik University has been supported by grant VEGA No. 1/0204/18, the grant of the Slovak Research and Development Agency under the contract No. APVV-17-0059. The work at MIT (RR JL YS) was supported in part by the Skoltech-MIT Center for Electrochemical Energy and the Cooperative Agreement between the Masdar Institute, Abu Dhabi, UAE and the MIT, Cambridge, MA, USA (02/MI/MIT/CP/11/07633/GEN/G/00). One of the authors (TK) thanks the Japanese Society for the Promotion of Science (JSPS) for JSPS Postdoctoral Fellowships for Research Abroad.